\documentclass[preprint]{aastex63}

\newcommand\rrle{RR Lyr\ae}
\newcommand\rrl{RRL}
\newcommand\rrls{RRLs}
\newcommand\wuma{W~UMa}

\defcitealias{Carretta2009}{C09}
\defcitealias{1997A&AS..121...95C}{CG97}
\defcitealias{ZinnWest1984}{ZW84}
\defcitealias{2021ApJ...908...20C}{C21}
\defcitealias{2007MNRAS.374.1421M}{M07}
\defcitealias{2014IAUS..301..461M}{M14}
\defcitealias{1996A&A...312..111J}{JK96}
\defcitealias{2013ApJ...773..181N}{N13}
\defcitealias{2020arXiv200802280I}{IB20}
\defcitealias{Mullen2021}{Paper I}
\defcitealias{2016MNRAS.461L..41M}{MV16b}
\defcitealias{2016MNRAS.462.4349M}{MV16a}

\received{December 8, 2021}
\revised{April 4, 2022}
\accepted{April 15, 2022}
\submitjournal{ApJ}

\shorttitle{RRc Fourier Metallicity from Light Curves}
\shortauthors{Mullen et al.}
\graphicspath{{./}{figures/}}

\begin{document}

\title{Metallicity of Galactic \rrle{} from Optical and Infrared Light Curves:\\ II. Period-Fourier-Metallicity Relations for First Overtone \rrle{}}

\correspondingauthor{Joseph P. Mullen}
\email{jpmullen@iastate.edu}

\author[0000-0002-1650-2764]{Joseph P. Mullen}
\affiliation{Department of Physics and Astronomy, Iowa State University, Ames, IA 50011, USA}

\author[0000-0001-9910-9230]{Massimo Marengo}
\affiliation{Department of Physics and Astronomy, Iowa State University, Ames, IA 50011, USA}

\author[0000-0002-9144-7726]{Clara E. Mart\'inez-V\'azquez}
\affiliation{Gemini Observatory, NSF's NOIRLab, 670 N. A'ohoku Place, Hilo, HI 96720, USA}
\affiliation{Cerro Tololo Inter--American Observatory, NSF's NOIRLab, Casilla 603, La Serena, Chile}

\author[0000-0002-4896-8841]{Giuseppe Bono}
\affiliation{Dipartimento di Fisica, Universit\`a di Roma Tor Vergata, via della Ricerca Scientifica 1, 00133 Roma, Italy}
\affiliation{INAF -- Osservatorio Astronomico di Roma, via Frascati 33, 00078 Monte Porzio Catone, Italy}

\author[0000-0001-7511-2830]{Vittorio F. Braga}
\affiliation{INAF -- Osservatorio Astronomico di Roma, via Frascati 33, 00078 Monte Porzio Catone, Italy}
\affiliation{Space Science Data Center -- ASI, via del Politecnico snc, 00133 Roma, Italy}

\author[0000-0003-3096-4161]{Brian Chaboyer}
\affiliation{Department of Physics and Astronomy, Dartmouth College, 6127 Wilder Laboratory, Hanover, NH 03755, USA}

\author[0000-0001-8926-3496]{Juliana Crestani}
\affiliation{Dipartimento di Fisica, Universit\`a di Roma Tor Vergata, via della Ricerca Scientifica 1, 00133 Roma, Italy}
\affiliation{INAF -- Osservatorio Astronomico di Roma, via Frascati 33, 00078 Monte Porzio Catone, Italy}
\affiliation{Departamento de Astronomia, Universidade Federal do Rio Grande do Sul, Av. Bento Gon\c{c}alves 6500, Porto Alegre 91501-970, Brazil}

\author[0000-0001-8209-0449]{Massimo Dall'Ora}
\affiliation{INAF -- Osservatorio Astronomico di Capodimonte, Salita Moiariello 16, 80131 Napoli, Italy}

\author[0000-0001-5829-111X]{Michele Fabrizio}
\affiliation{INAF -- Osservatorio Astronomico di Roma, via Frascati 33, 00078 Monte Porzio Catone, Italy}
\affiliation{Space Science Data Center -- ASI, via del Politecnico snc, 00133 Roma, Italy}

\author[0000-0003-0376-6928]{Giuliana Fiorentino}
\affiliation{INAF -- Osservatorio Astronomico di Roma, via Frascati 33, 00078 Monte Porzio Catone, Italy}

\author[0000-0001-5292-6380]{Matteo Monelli}
\affiliation{IAC- Instituto de Astrof\'isica de Canarias Calle V\'ia Lactea s/n, E-38205 La Laguna, Tenerife, Spain\\}
\affiliation{Departmento de Astrof\'isica, Universidad de La Laguna, E-38206 La Laguna, Tenerife, Spain\\}

\author[0000-0002-8894-836X]{Jillian R. Neeley}
\affiliation{Department of Physics, Florida Atlantic University, 777 Glades Rd, Boca Raton, FL 33431}

\author[0000-0001-6074-6830]{Peter B. Stetson}
\affiliation{Herzberg Astronomy and Astrophysics, National Research Council, 5071 West Saanich Road, Victoria, British Columbia V9E 2E7, Canada}

\author[0000-0002-5032-2476]{Fr\'ed\'eric Th\'evenin}
\affiliation{Universit\'e de Nice Sophia--antipolis, CNRS, Observatoire de la C\^ote d'Azur, Laboratoire Lagrange, BP 4229, F-06304 Nice, France}



\begin{abstract}
We present new period-$\phi_{31}$-[Fe/H] relations for first overtone \rrl{} stars (RRc), calibrated over a broad range of metallicities ($-2.5 \la \textrm{[Fe/H]}\la 0.0$) utilizing the largest currently available set of Galactic halo field \rrl{} with homogeneous spectroscopic metallicities. Our relations are defined in the optical (ASAS-SN $V$-band) and, inaugurally, in the infrared (WISE $W1$ and $W2$ bands). Our $V$-band relation can reproduce individual RRc spectroscopic metallicities with a dispersion of 0.30~dex over the entire metallicity range of our calibrator sample (an RMS smaller than what we found for other relations in literature including non-linear terms). Our infrared relation has a similar dispersion in the low and intermediate metallicity range ($\textrm{[Fe/H]} \la -0.5$) but tends to underestimate the [Fe/H] abundance around solar metallicity. We tested our relations by measuring both the metallicity of the Sculptor dSph and a sample of Galactic globular clusters, rich in both RRc and RRab stars. The average metallicity we obtain for the combined \rrl{} sample in each cluster is within $\pm 0.08$~dex of their spectroscopic metallicities. The infrared and optical relations presented in this work will enable deriving reliable photometric RRL metallicities in conditions where spectroscopic measurements are not feasible; e.g., in distant galaxies or reddened regions (observed with upcoming Extremely Large Telescopes and the James Webb Space Telescope), or in the large sample of new \rrl{} that will be discovered in large-area time-domain photometric surveys (such as LSST and the Roman space telescope).

\end{abstract}



\section{Introduction} \label{sec:intro}

RR~Lyr\ae{} stars (\rrl{} hereafter) are a ubiquitous and widely used tracer of old (age $> 10$~Gyr, \citealt{Walker1989,2020A&A...641A..96S}) stellar populations in the Milky Way (MW)  and Local Group (LG) galaxies (see e.g., MW: \citet{2021arXiv210613145M}, Carina: \citet{2015ApJ...814...71C}; Sculptor: \citet{2016MNRAS.462.4349M}; M31 dwarf satellites: \citet{2017ApJ...850..137M,2017ApJ...842...60M}; Isolated LG dwarfs: \citet{2009ApJ...699.1742B,2010ApJ...712.1259B,2013MNRAS.432.3047B}). Their relevance has become even more important in the current age of large area photometric time surveys (e.g., ASAS-SN \citep{2014ApJ...788...48S, 2018MNRAS.477.3145J}, Catalina Sky Survey \citep{2009ApJ...696..870D}, PanSTARRS \citep{2016arXiv161205560C}, DES \citep{2018ApJS..239...18A}, Gaia \citep{2016A&A...595A.133C,2019A&A...622A..60C}, TESS \citep{2015JATIS...1a4003R}), which have significantly increased the number of known \rrl{} variables. An even larger number of these variables is expected to be discovered as the product of upcoming next-generation surveys, such as the Vera C. Rubin Observatory Legacy Survey of Space and Time \citep[LSST,][]{2019ApJ...873..111I} in the optical, and surveys that will be executed for the Nancy Grace Roman telescope \citep{2019arXiv190205569A} at near-infrared wavelength.

The importance of \rrl{} as tracers is related to their role as distance indicators. Period-Wesenheit-Metallicity relations (PWZ) in the optical and Period-Luminosity-Metallicity relations (PLZ) in the infrared (theoretical: \citet{2015ApJ...808...50M, 2017ApJ...841...84N} and observational: \citet{2013MNRAS.435.3206D, 2018MNRAS.481.1195M, 2019MNRAS.490.4254N,2021MNRAS.503.4719G}) now provide individual \rrl{} distances with an accuracy approaching other traditional stellar standard candles that can be characterized by a Leavitt Law \citep{1908AnHar..60...87L,1912HarCi.173....1L}, such as Cepheids. 

In the case of Classical (Population I) Cepheids, the metallicity dependence on luminosity may, in most cases, be ignored (but see e.g., \citealt{2021ApJ...913...38B}); however, the much larger spread in metallicity for \rrl{} (Population II stars) requires reliable measurements of their [Fe/H] abundance to provide accurate distances. Having metallicity measurements available can be a challenge since they are traditionally derived with spectroscopic methods, and spectra cannot be expected to be readily available for the large sample of variables being discovered in large-area surveys. This can be due to logistical constraints based upon telescope time or physical limitations for taking spectra such as extreme distances or high extinction in environments that can only be probed photometrically by mid-infrared cameras. A reliable and precise method to extract physical parameters based solely on photometric time series is necessary to keep pace with this expansion.

\citet{1981ApJ...248..291S} first showed for variable stars that specific physical parameters, such as metallicity, could be directly related to a light curve's shape (characterized through its Fourier decomposition). \citet{1996A&A...312..111J} quantified this relationship by finding a bi-linear relation between period and a low order Fourier phase parameter ($\phi_{31} = \phi_3 - 3 \cdot \phi_1$) derived in the optical, for \rrls{} pulsating in the fundamental mode (also know as RRab). Additional works have analyzed and revised this relation for RRab (\citealt{2013ApJ...773..181N,Martinez-Vazquez2016,2005AcA....55...59S,2016ApJS..227...30N,2020arXiv200802280I,Mullen2021}) using more modern datasets in a variety of wavelengths. In particular, \citealt{Mullen2021} (hereafter \citetalias{Mullen2021}) obtained a new relation in the $V$-band based on the largest RRab calibration dataset to date. In \citetalias{Mullen2021}, we showed that accurate RRab photometric metallicities could be extended to both lower and higher metallicity regimes than in previous works. Furthermore, for the first time we showed that a photometric metallicity relation for RRab variables could be extended into the mid-infrared. This paper aims to now derive similar relations for \rrl{} pulsating in the first overtone (RRc variables).

The study of RRc variables possesses additional challenges with respect to RRab. An important issue is the sample size, as the number of RRc is only $\sim 1/3$ of all field \rrl{} (the ratio of RRc to the total number of \rrl{} is intimately tied to how [Fe/H] affects horizontal branch morphology; \citealt{2021arXiv210700919F}). Furthermore, extracting accurate periods and Fourier parameters for RRc often takes an extra layer of scrutiny stemming from their inherent smaller amplitudes and quasi-sinusoidal light curves, when compared to the characteristic high-amplitude saw-toothed shape RRab which are more strongly dependent on atmospheric abundances. Due in part to these issues, it was not until \citet{2007MNRAS.374.1421M} (hereafter M07) that it was shown that similar period-$\phi_{31}$-[Fe/H] equations could be made for RRc in the $V$-band. Further work by \citet{2013ApJ...773..181N} (hereafter N13), \citet{2014IAUS..301..461M} (hereafter M14), and \citet{2020arXiv200802280I} (hereafter IB20) extended this analysis, respectively, to well-sampled RRc light curves obtained with the Kepler space telescope \citep{2010ApJ...713L..79K}, to a revised sample of globular clusters (GC) with $V$-band photometry, and by using Gaia DR2 \citep{2018A&A...616A...1G,2018A&A...618A..30H,2019A&A...622A..60C} $G$-band light curves.

Due to the scarcity of accurate high-resolution ($R \ga 20$,000) spectroscopic metallicity measurements  (capable of providing accuracy of $\sim 0.1$~dex in [Fe/H]) for individual RRc, these relations have all been predominantly based on RRc residing in globular clusters (GCs) with well-studied cluster metallicity. In this work, we leverage newly determined RRc high-resolution (HR) metallicities from \citet{2021ApJ...908...20C} (\citetalias{2021ApJ...908...20C} hereafter). We combine these measurements with a large sample of metallicities derived with medium-resolution spectroscopic surveys ($R \sim 2$,000). This new set of medium-resolution metallicities has been derived with the updated \citetalias{2021ApJ...908...20C} calibration of the $\Delta S$ method \citep{1959ApJ...130..507P}, which relies on ratios between the equivalent widths of Ca and H lines. While not as accurate as HR metallicities, this method can nevertheless provide reliable [Fe/H] abundances with a typical uncertainty of 0.2-0.3~dex.

This work directly follows the analysis presented in \citetalias{Mullen2021} for the calibration of period-$\phi_{31}$-[Fe/H] relations for RRab field stars. We take advantage of an extensive catalog of field RRc [Fe/H] abundances from \citetalias{2021ApJ...908...20C} HR spectral measurements, combined with $\Delta S$ metallicities estimated from large publicly available medium-resolution spectral datasets. We then  cross-correlate this extensive HR+$\Delta$S metallicity catalog with well-sampled archival light curves in order to derive novel period-$\phi_{31}$-[Fe/H] relations for RRc in the optical ($V$-band) and, for the first time, mid-infrared ($W1$ and $W2$ bands). Our work shows these relations can indeed be extended to the thermal infrared, where the light curves are determined mainly by the radius variation during the star's pulsation rather than the effective temperature changes that dominate in the optical wavelengths. This will be crucial to allow the determination of reliable metallicities in upcoming space infrared surveys, such as the James Webb Space Telescope \citep[JWST,][]{2006SSRv..123..485G}.

This paper is structured as follows. In Section~\ref{sec:datasets}, we describe the data sets we adopt for our work: the HR+$\Delta$S metallicity catalog utilizing the work of \citetalias{2021ApJ...908...20C} and the optical and infrared time-series catalogs from which we derived the RRc light curves. In Section~\ref{sec:calibration}, we explain how our period-$\phi_{31}$-[Fe/H] relations are calibrated. We then validate our sample to remove contaminants such as eclipsing contact binaries which possess similar light curves to RRc and can often be missclassified. Our results are discussed in Section~\ref{sec:discussion}, where we assess the precision of the optical and infrared relations, compare our relations with previous ones found in literature, and apply our method to measure the [Fe/H] abundance in a sample of globular clusters. Lastly, we derive the metallicity distribution of the Milky Way's dwarf spheroidal (dSph) satellite Sculptor and compare it to the metallicities available using other solely photometric methods.
Our conclusions are presented in Section~\ref{sec:conclusions}.


\section{First Overtone RR Lyrae Datasets} \label{sec:datasets}

Our sample of \rrls{} is derived from an extensive catalog of 3057 field RRc, for which we have either [Fe/H] abundances derived from HR spectra (40 sources) or an estimate of their metallicity based on the $\Delta S$ method (3017 sources). In order to ensure a homogeneous metallicity scale for our entire sample, both HR and $\Delta S$ metallicities are based on the calibration provided in \citetalias{2021ApJ...908...20C} (in turn consistent with the \citealt{Carretta2009} metallicity scale). The RRc stars for which we obtained $\Delta S$ metallicities were extracted from both the full Large Scale Area Multi-Object Spectroscopic Telescope (LAMOST) DR2 survey \citep{2012RAA....12..735D, 2014IAUS..298..310L} and the Sloan Extension for Galactic Understanding and Exploration \citep[SEGUE,][]{2009AJ....137.4377Y} survey datasets. For a complete and detailed description of the metallicity scale, the HR metallicity catalog's demographics, the $\Delta$S calibration, and the spectrum selection criteria, we refer the reader to the \citetalias{2021ApJ...908...20C} paper.

We have then cross-matched the variables in our HR+$\Delta$S metallicity catalog with well-sampled photometric time series in the All-Sky Automated Survey for Supernovae (ASAS-SN,  \citealt{2014ApJ...788...48S, 2018MNRAS.477.3145J}) and the Near-Earth Objects reactivation mission (NEOWISE, \citealt{2011ApJ...731...53M}) of the  Wide-field Infrared Survey Explorer (WISE, \citealt{2010AJ....140.1868W}). From the ASAS-SN survey, we extracted 594 good quality ASAS-SN RRc light curves in the $V$-band. Similarly, from the WISE/NEOWISE missions we obtained 106 good quality infrared light curves in at least one of the available bands. Individually, the $W1$ (3.4~$\micron$) and $W2$ (4.6~$\micron$) bands had 106 and 71 good quality RRc stellar light curves respectively. From hereafter in the paper, the combination of photometry from both the primary WISE and ongoing NEOWISE mission will be referred to for brevity as WISE. Table~\ref{tab:calibrators} lists the properties of the RRc in our calibration samples that have passed the stringent photometric and Fourier decomposition criteria described in Section~\ref{sec:calibration}. The \emph{Joint Sample} column denotes the subset of 83 RRc for which both optical and infrared light curves are available.

\begin{table}[!h]

    \begin{center}
    \caption{Calibration Datasets}\label{tab:calibrators}
    \begin{tabular}{lcccc}
    \tableline
    & ASAS-SN & \multicolumn{2}{c}{WISE}& Joint Sample \\
    \tableline
    \tableline
    Bands & $V$ & $W1$ & $W2$ & $V$, ($W1$ or $W2$)  \\
    RRc stars & 594 & 106 & 71 & 83 \\
    Period in days (range) & 0.21 - 0.43 & 0.23 - 0.41 & 0.24 - 0.41 & 0.23 - 0.41 \\
    Period in days (median value)& 0.32 & 0.32 & 0.32 & 0.33 \\
    $[$Fe/H$]$ (range) & $-$2.54 - ($-$0.16) & $-$2.54 - (+0.55)  & $-$2.54 - (+0.06) & $-$2.54 - ($-$0.16) \\
    $[$Fe/H$]$ (median value) & $-$1.60 & $-$1.64 & $-$1.65 & $-$1.66 \\
    Number of epochs (range)\tablenotemark{a} & 118 - 835 & 159 - 2942 & 159 - 2942 & \nodata \\
    Number of epochs (median value) & 230 & 205 & 212 & \nodata \\
    Magnitude (range) & 9.12 - 16.63 & 8.10 - 13.01 & 8.12 - 12.75 & \nodata \\ 
    \tableline
    \end{tabular}  
    \tablenotetext{a}{The number of epochs is recorded prior to removing any spurious photometric measurement (see Section~\ref{ssec:processing}).}
    \end{center}

\end{table}

\begin{figure}[!h]
    \centering
    \includegraphics[width=\textwidth]{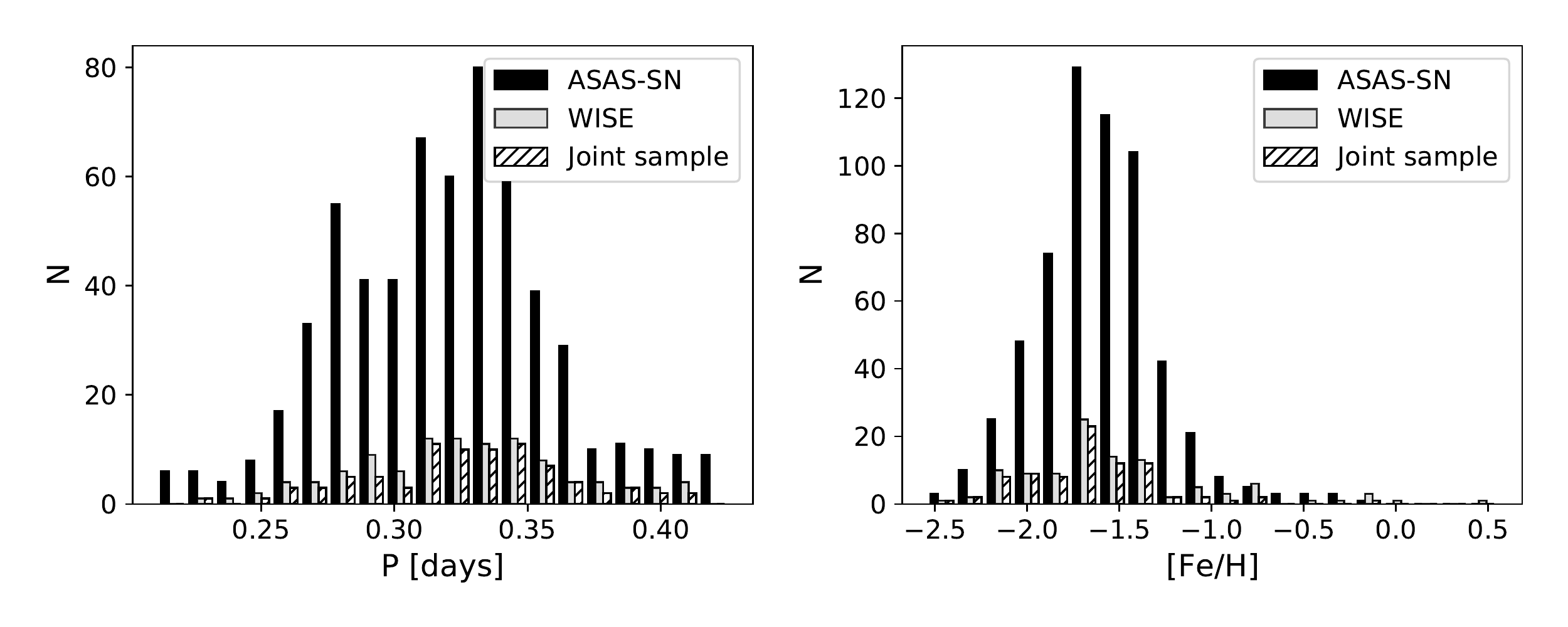}
    \caption{Period distribution (left) and spectroscopic [Fe/H] distribution (right) of the different calibration datasets. The histogram labeled ``Joint sample'' (hatched) corresponds to those stars in common between the ASAS-SN (optical $V$-band, in black) and WISE (infrared $W1$ and/or $W2$ bands, in grey) datasets.}
    \label{fig:P&Feh}
\end{figure}

Figure~\ref{fig:P&Feh} shows the distribution of period and metallicity for the optical (ASAS-SN) and infrared (WISE) calibration samples and the stars in common (joint). All samples cover the entire period range expected for RRc variables and are representative of the metallicity of Galactic Halo \rrls{}, with a median [Fe/H] abundance of $\approx -1.6$ in both the ASAS-SN and WISE samples. The broad range of metallicities apparent in Figure~\ref{fig:P&Feh} ensures ample leverage for calibrating our period-$\phi_{31}$-[Fe/H] relations.

\begin{center}
\begin{figure}[!h]
       \includegraphics[width=\textwidth]{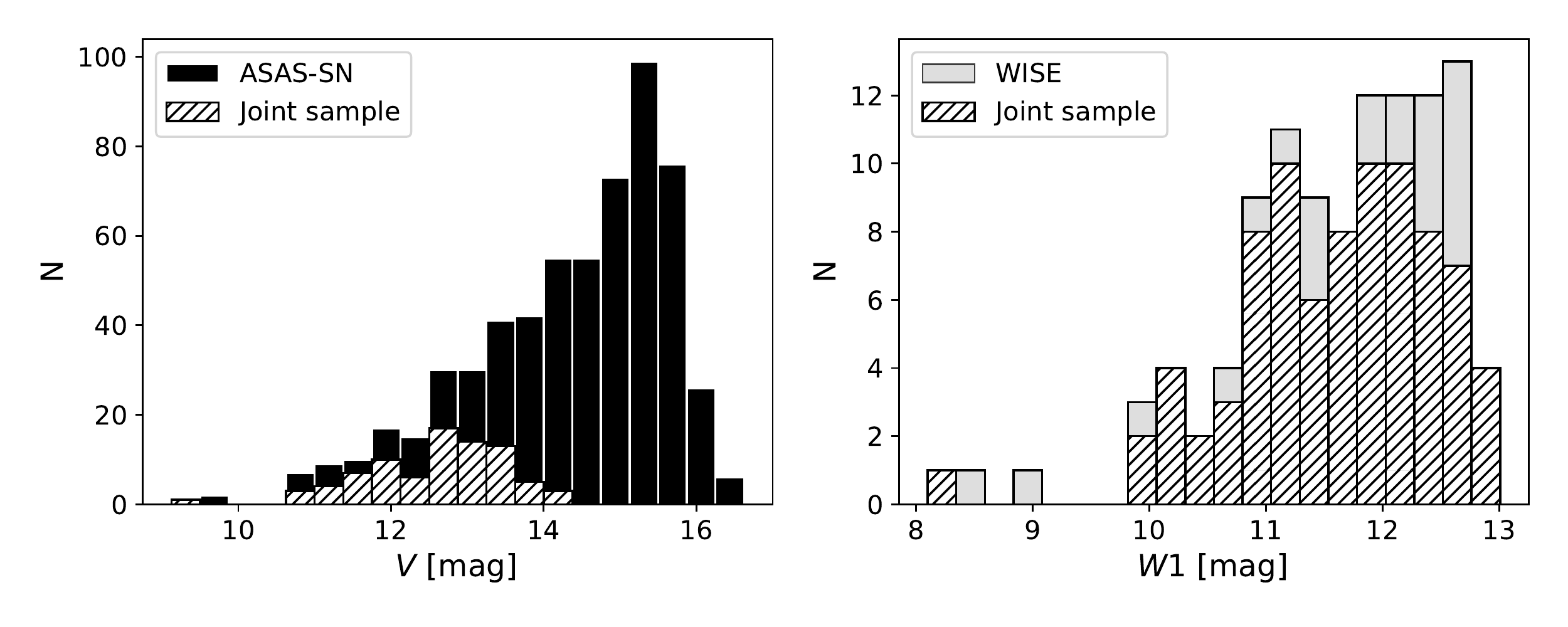}
    \caption{Distribution of the average $V$-band apparent magnitude (left) and the average $W1$-band apparent magnitude (right). Each panel shows both the entire calibration sample (solid fill) and the joint sample (hatched fill, subset of stars having both ASAS-SN and WISE data).}
    \label{fig:mag_distrib}
\end{figure}
\end{center}

The apparent magnitude distribution of the calibration datasets is shown in Figure~\ref{fig:mag_distrib} for the $V$-band (left panel) and $W1$-band (right). The histogram for the $W2$-band is similar, although fewer in stars due to the lower sensitivity in this band, resulting in generally noisier light curves. The joint sample (hatched in both panels) is noticeably truncated at $V\sim$~14 mag due to the WISE survey's shallower photometric depth.

\section{Calibration of Period-Fourier-Metallicity relation}\label{sec:calibration}

\subsection{Data Processing}\label{ssec:processing}

\begin{figure}[!h]
    \centering
    \includegraphics[width=\textwidth]{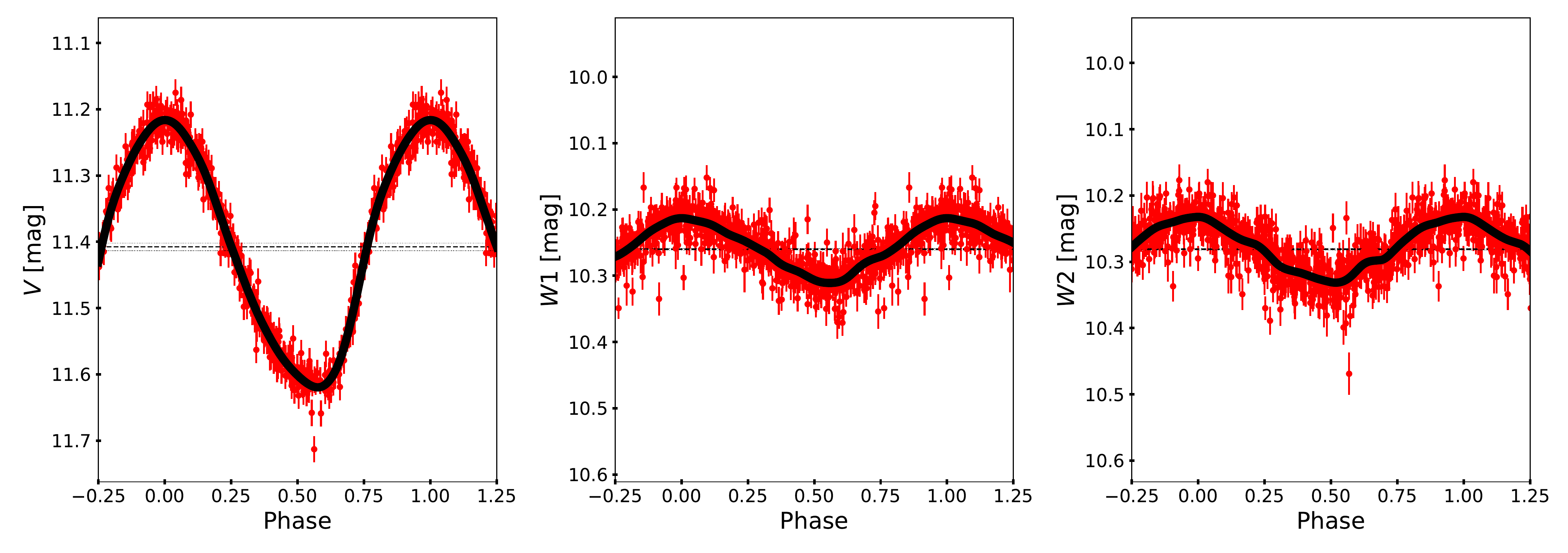}
    \caption{Multiple-band light curves for a typical RRc star. The $V$ (left), $W1$ (middle), and $W2$ bands (right) are shown for the star XX Dor (Period = 0.32894 days). The Fourier fit (black solid line) to the GLOESS light curve (see text) is plotted on top of the phased data (red). The average magnitude with its associated error are shown as horizontal black dotted lines. Points automatically rejected by the GLOESS fitting procedure have been removed.}
    \label{fig:LCV}
\end{figure}

The extraction of period and Fourier parameters directly follows the procedure outlined in \citetalias{Mullen2021}, Section 3, and is summarized below. We first refine the period of each variable with the Lomb-Scargle method \citep{1976Ap&SS..39..447L, 1982ApJ...263..835S}, applied to the large temporal baseline ($>8$ years) of the ASAS-SN and WISE time series data. A Gaussian locally-weighted regression smoothing algorithm (GLOESS, \citealt{2004AJ....128.2239P}) is then applied to the phased data in order to smooth over unevenly sampled data and exclude outliers. A Fourier sine or cosine decomposition is finally executed on each GLOESS light curve by applying a weighted least-squares fit of the following equation (in the case of the sine decomposition): 

\begin{equation}\label{eq:fourier_expansion}
        m(\Phi)=A_0+\sum_{i=1}^{n} A_{i}\sin[2\pi i  (\Phi+\Phi_{0})+ \phi_{i}]
\end{equation}

\noindent
where $m(\Phi)$ is the observed magnitude for either the ASAS-SN or WISE bands, $A_{0}$ is the mean magnitude, $n$ is the order of the expansion, $\Phi$ is the phase from the GLOESS light curve varying from 0 to 1, $\Phi_{0}$ is the phase that corresponds to the time of maximum light $T_{0}$, $A_{i}$ and $\phi_{i}$ are the $i$-th order Fourier amplitude and phase coefficients, respectively. A similar equation can be written to represent the equivalent cosine decomposition. For RRc n is often less than five; however, we advise the reader to carefully choose the optimal number of terms for their dataset (see \citealt{1986A&A...170...59P,2009A&A...507.1729D}) as the addition of higher-order terms may cause inconsistencies in the value of low order Fourier parameters. For this work, we note that by using an intermediate GLOESS step to eliminate noise factors, the Fourier parameters are consistent between 3, 4, or fifth order decompositions, with the exclusion of a small amount of $\gg 3\sigma$ outliers ($\sim4\%$ of total RRc for ASAS-SN and $\sim8\%$ for WISE). This small subset of outliers has been excluded in our processing to make our result invariant between different decomposition orders used. However, it is worth noting that with the robust fitting procedure outlined in Section \ref{ssec:planefit}, the results of this paper remain unaltered with the inclusion/exclusion of these sources. Figure~\ref{fig:LCV} shows a typical $V$, $W1$, and $W2$ light curve from our RRc calibration stars where the Fourier decomposition fit (solid black line) is plotted on top of the actual photometric data in red.

\begin{figure}[!h]
    \centering
    \includegraphics[width=\textwidth]{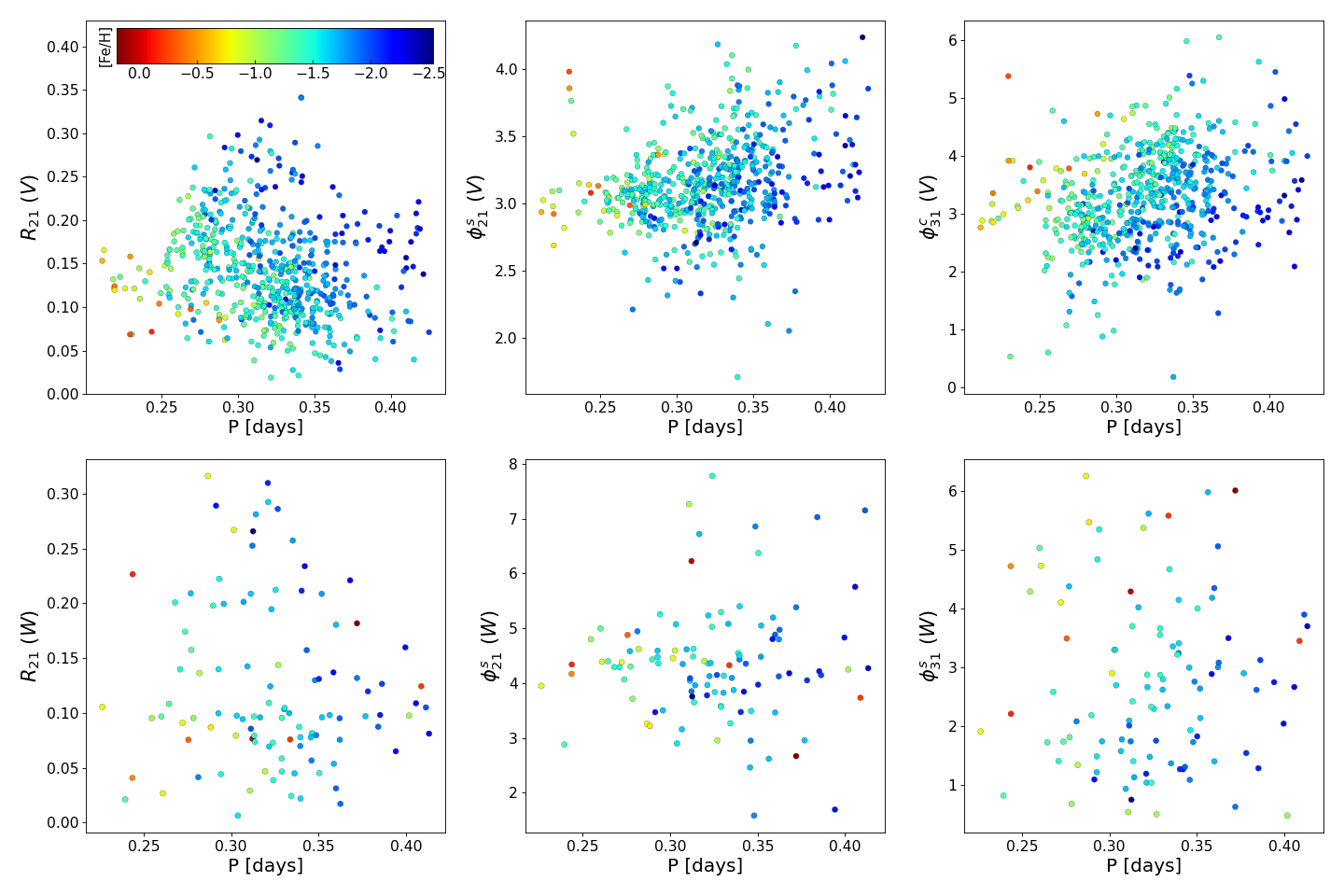}
    \caption{ASAS-SN $V$-band (top row) and $W1$-band (bottom row) Fourier parameters as functions of the period. \rrls~are color-coded based on their spectroscopic calibration metallicities. In the rightmost column ($\phi_{31}$), [Fe/H] generally goes from metal-rich in the left (red) to metal-poor in the right (blue). All parameters have been derived with a sine Fourier decomposition (indicated with a $s$ superscript), except for the top-right panel ($\phi_{31}^c (V)$ values) where the Fourier phase parameters have been converted to a cosine decomposition (see text). Average errors for each of the parameters plotted are smaller than the size of the data points.}
    \label{fig:FOURIER_COEFFS}
\end{figure}

As in \citetalias{Mullen2021}, we explore the link between metallicity and Fourier coefficients through either the ratios of Fourier amplitudes $R_{ij}=A_{i}/A_{j}$, or the linear combinations of the phase coefficients $\phi_{ij}=j\cdot \phi_{i}-i \cdot\phi_{j}$, where $\phi_{ij}$ is cyclic in nature and ranges from $0$ to $2\pi$. This is shown in Figure~\ref{fig:FOURIER_COEFFS}, where several low-order Fourier parameters of individual RRc stars are plotted against period and color-coded based on their [Fe/H]. In the case of the RRc $V$-band $\phi_{31}$ values (top-right), it is advantageous to represent the phase parameter as a product of a cosine decomposition rather than sine to avoid the rollover of the $\phi_{31}$ parameter across the $2\pi$ boundary. This can be achieved by either adopting the cosine form of equation~\ref{eq:fourier_expansion} or using the simple transformation $\phi_{31}^c=\phi_{31}^s - \pi$ between the sine ($s$ superscript) and cosine ($c$ superscript) form of this parameter.

In all $V$-band panels, we can readily see a gradient in the [Fe/H] distribution of the stars. We found the most distinct separation to be in the $\phi_{31}^c$ vs. period plot, confirming earlier studies, such as \citetalias{2007MNRAS.374.1421M}, which suggests a relation between period and $\phi_{31}^c$ to be a good indicator of metallicity for the $V$-band RRc. We similarly found that the $\phi_{31}^s$ vs. period plot for WISE (bottom-right panel) showed the most distinct trend with metallicity. However, the general metallicity trends in the WISE bands are less apparent than in the optical due to both the smaller sample size and the larger intrinsic dispersion of $\phi_{31}$ values, covering almost the entire 2$\pi$ range. As in the case of the RRab stars, we take advantage of the nearly identical shape of the light curves in the $W1$ and $W2$ bands (see e.g., Figure~\ref{fig:LCV}) to improve the signal in the WISE Fourier parameters, by averaging the $\phi_{31}^s (W1)$ and $\phi_{31}^s (W2)$ values for stars with both light curves available. See \citetalias{Mullen2021}, Section~3.2 for a quantitative discussion demonstrating the validity of this approach for the RRab stars; we have verified this result still holds for our RRc sample.

\subsection{Eclipsing Contact Binary Contamination}\label{ssec:Binary}

Upon initial creation of our $\Delta$S metallicity sample, we found a significant population of stars that were previously marked as RRc in literature but, on further analysis, were found to be likely eclipsing contact binaries known as W Ursae Majoris (\wuma{}). Those binaries have sinusoidal light curves similar to RRc variables, and can be confused with first overtone \rrl{} stars if they happen to have similar period and amplitude. If the noise is high enough in a \wuma{} light curve, it can prevent detecting a characteristic secondary eclipse.

\begin{figure}[!h]
    \centering
    \includegraphics[width=\textwidth]{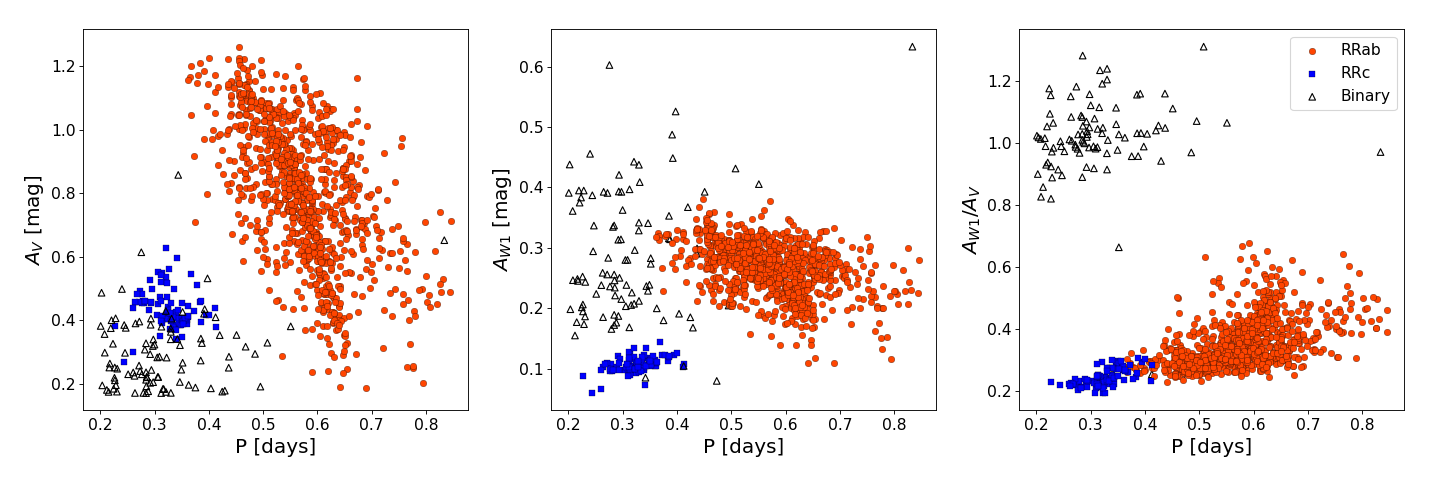}
    \caption{Period versus amplitude diagram for the ASAS-SN $V$-band (left) and the WISE $W1$-band (middle). The right panel shows the ratio of the $W1$ to $V$ amplitudes. Data plotted represents the subset of stars where both optical and infrared are available, i.e. the RRc joint sample discussed in Section~\ref{sec:datasets} (blue squares), the RRab joint sample (red circles) from \citetalias{Mullen2021}, and eclipsing contact binary contaminants selected as explained in Section~\ref{ssec:Binary} (empty triangles).}
    \label{fig:binary}
\end{figure}

Indeed, we found that for our sample of \wuma{}, most combinations of Fourier parameters show some degree of degeneracy with regions typical of RRc. In particular, Figure~\ref{fig:binary} shows that some \wuma{} binaries largely overlap with RRc variables in the optical Bailey diagram, although their amplitude tends to be \emph{smaller}. The overlap is partially mitigated in the infrared ($W1$, shown in figure, or $W2$ diagrams), in which case the same \wuma{} tends to have \emph{larger} amplitudes than RRc with a similar period. This opposite trend of amplitudes with wavelengths provided us with a simple criterion to automatically identify and reject the \wuma{} binaries from our sample in the case both bands are available (as in our joint sample): as shown in the left panel of Figure~\ref{fig:binary} the $A_{W1}/A_V$ ratio provides a clear separation between RRc stars and the potential binaries.

For all sources with only one available light curve (either in the optical or infrared), we resorted to less reliable criteria and manual inspection of the light curves. These criteria were developed for individual bands by analyzing an initial sample of the potential \wuma{} that have been identified with their amplitude ratio between infrared and optical. If only the $V$-band photometry was available, we flagged as potential \wuma{} all sources with $A_{V} \la 0.15$ or $\phi_{21}^s (V)>4.25$. For stars only having infrared data, we instead flagged all stars with $A_{W}>0.15$ (in either the $W1$ or $W2$ bands, depending on availability). We then visually examined the phased light curve in the one available band, searching for a secondary light curve minimum, which could have been missed in previous work based on less detailed photometry. We then rejected all sources showing evidence of such secondary eclipses.

The analysis presented in this paper (including Table~\ref{tab:calibrators}) focuses solely on this sample that has been cleaned of \wuma{}. We provide this final cautionary note that certain \wuma{} as shown can be easily misconstrued as RRc, and anyone using the relations published in this work should take extreme care in the validity of their RRc sample.

\subsection{Period-Fourier-[Fe/H] Fitting} \label{ssec:planefit}

As discussed in Section~\ref{sec:intro}, RRc photometric metallicities based on a Fourier decomposition of optical light curves are available for different literature relations (see e.g. \citetalias{2007MNRAS.374.1421M}, \citetalias{2013ApJ...773..181N}, \citetalias{2014IAUS..301..461M}, and \citetalias{2020arXiv200802280I}). These relations are either linear in both period and $\phi_{31}$ or use higher-order combinations of these two parameters. Our tests show that, for our sample, higher-order and non-linear terms in the period, $\phi_{31}$, or other Fourier parameters result in minimal benefits at the expense of complexity and decreased robustness in the fitting results. For these reasons, we decided to proceed using a simple linear relation in period and $\phi_{31}$, similarly to what we did in \citetalias{Mullen2021} for the RRab stars.

To validate this choice of a period-$\phi_{31}$ plane-fit, we followed the methodology described in Appendix B of \citetalias{Mullen2021}. Principal component analysis showed that 91.30\% of the variance in the $V$-band RRc data could be attributed to just two principal axes (i.e., a plane). This is comparable to the 92.9\% value we found for RRab stars in \citetalias{Mullen2021}, which we interpreted as suggesting that a third fit parameter was not needed. A similar analysis for the infrared RRc data set showed that 85.17\% of the variance could be attributed to two dimensions (6.74\% less than what we found for the RRab in the WISE bands). We attribute this larger variance to a larger relative scatter in the WISE dataset itself (seen in the bottom-right panel of Figure~\ref{fig:FOURIER_COEFFS}) due to the smaller amplitude of the pulsations, higher intrinsic photometric noise in the data, and smaller size of the sample. Search for a third observable parameter and/or higher-order fitting functions did not result in a better fit, leading us to adopt a period-$\phi_{31}$ plane-fit also for the infrared dataset.

We initially attempted to fit an RRc period-$\phi_{31}$-[Fe/H] relation adopting the same procedure described in \citetalias{Mullen2021}. We quickly discovered that the smaller size of the RRc sample (by a factor of 3 and 10, compared to the RRab, in the ASAS-SN and WISE bands, respectively) resulted in a fit which was highly dependent upon the exact calibration sample utilized. Bootstrap analysis \citep{10.1214/ss/1177013815} highlighted that this instability in the fitting results was due to a combination of higher susceptibility to individual outliers and a stronger correlation between the period and $\phi_{31}$ slopes. We, however, found that when utilizing different bootstrap samples, the range of period slopes we were obtaining for the best fit RRc relation was consistent with the value we derived in \citetalias{Mullen2021} for the RRab relation. Based on these considerations, we decided to freeze the period slope of the RRc period-$\phi_{31}$-[Fe/H] relation to the RRab value, effectively reducing the RRc fit to a two-parameter fit in the metallicity zero point $a$ and the $\phi_{31}$ slope $c$:

\begin{equation}\label{eq:plane}
    \textrm{[Fe/H]} = a + b_{F} \cdot (P_{F} - P_0) + c \cdot(\phi_{31} - \phi_{31_0})
\end{equation}

\noindent
where $b_F$ is the period slope from the fit of RRab stars in \citetalias{Mullen2021}, equal to $-7.60 \pm 0.24$~dex/day for the ASAS-SN $V$-band, and $-8.33 \pm 0.34$~dex/day for the WISE bands. Note that, for this choice of values to be appropriate, the period of the RRc variables needs first to be ``fundamentalized'', or transformed into its equivalent fundamental period by using the equation $\log P_{F}=\log P_{FO}+0.127$ \citep{1971A&A....14..293I,1973ApJ...184..815R,1983ApJ...266...94C}. Fundamentalizing the period of overtone pulsators is a common technique when deriving PLZ and PWZ relations of Cepheids and \rrl{} variables (see, e.g, \citealt{2018A&A...619A...8G}, \citealt{2015ApJ...808...50M}, \citealt{2015ApJ...808...11N}) when their smaller number and narrower period range makes it prohibitive to fit them alone. It is worth noting that fundamentalization does have a minor dependence upon metallicity (\citealt{2001AJ....122..207B,2014AcA....64..177S}); however, the effect is so small that no matter the relation tried (e.g., \citealt{2015ApJ...814...71C}), the results of this paper remain unchanged. Finally, as in \citetalias{Mullen2021}, we have included the pivot offsets $P_0$ and $\phi_{31_0}$ to add further robustness to the fitting procedure; each value is close to the median of the period and [Fe/H] distribution for the RRc sample.

We fit the two remaining free parameters of Equation~\ref{eq:plane}, and their uncertainties, using an Orthogonal Distance Regression (ODR) routine combined with Bootstrap re-sampling. The analysis we presented in \citetalias{Mullen2021} for RRab stars showed that ODR tends to minimize trends along all fitted dimensions as no differentiation is made between dependent and independent variables. The ODR fit was run multiple times using bootstrap re-sampling, where in each run all the parameters of each individual calibrator star (period, $\phi_{31}$, and [Fe/H]) are randomly replaced with all the corresponding parameters from any singular star in the same calibration dataset. Namely, we are sampling with replacement our entire calibration dataset before fitting our relation, and ensuring each re-sampled calibration dataset is the same size as our initial sample. Note, all the stars in our calibration sample are fitted with equal weight, as the uncertainties in the fit are dominated by the intrinsic scatter we observe in the period-$\phi_{31}$-[Fe/H] plane (see Figure~\ref{fig:FOURIER_COEFFS}), in comparison to the small uncertainties in period and the $\phi_{31}$ parameter (on average $\sim 10^{-6}$ days and $\sim$0.02 radians respectively).

\begin{figure}[h!]
    \centering
    \includegraphics[width=\textwidth]{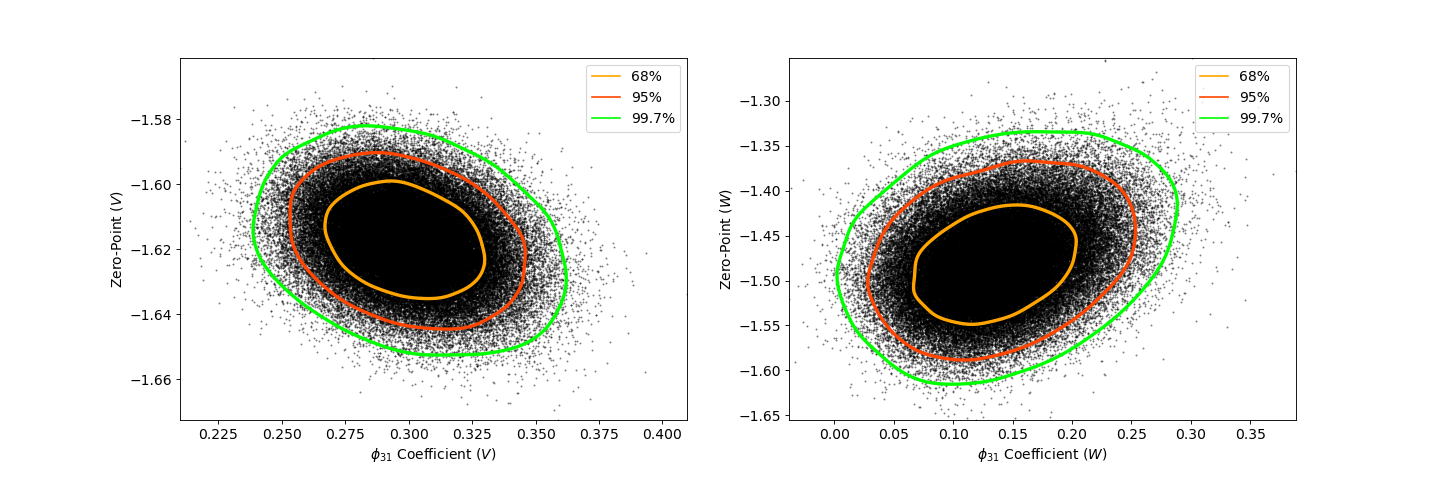}
    \caption{Fitting parameters from the bootstrap re-sampled ODR method applied to the ASAS-SN ($V$) sample left and WISE ($W$) sample right. Contours shown denote regions containing $68\%$ (red), $95\%$ (orange), and $99.7\%$ (green) of sampled parameters.}
    \label{fig:BOOTSTRAP}
\end{figure}

Figure~\ref{fig:BOOTSTRAP} shows the ``cloud'' of best fit values obtained with the process described above in the optical and infrared fit. The contours enclose the regions with 68\%, 95\%, and 99.7\% of the ODR best fit values obtained with 100,000 bootstrap re-sampled datasets. The contours provide a visual representation of the robustness of the fitting parameters with respect to outliers and are significantly larger than the corresponding 1, 2 and 3$\sigma$ error ellipses of the best-fit parameters of individual ODR fits. For this reason, we adopted the mean values and standard deviations of the bootstrap re-sampled ODR fits as solutions and uncertainties for our RRc period-$\phi_{31}$-[Fe/H] relations:

\begin{eqnarray}
    &&\textrm{[Fe/H]}_{V} = (-1.62 \pm 0.01)+(-7.60) \cdot (P_{F}-0.43)+(0.30 \pm 0.02) \cdot (\phi_{31}^c-3.20) \label{eq:fourier_ASASSN} \\
    &&\textrm{[Fe/H]}_{W} = (-1.48 \pm 0.05)+(-8.33) \cdot (P_{F}-0.43)+(0.14 \pm 0.05) \cdot (\phi_{31}^s-2.70) \label{eq:fourier_NEO}
\end{eqnarray}

\noindent
where Equations~\ref{eq:fourier_ASASSN} and \ref{eq:fourier_NEO} provide photometric metallicities for the ASAS-SN $V$-band and WISE $W1$ and $W2$ infrared bands (averaged when possible, see Section~\ref{ssec:processing}). Note again how the $V$-band metallicities are expressed as a function of the cosine decomposition of the optical lightcurves (i.e. they depend on $\phi_{31}^c$), while the WISE formula is a function of the sine decomposition parameter $\phi_{31}^s$.

Figure~\ref{fig:BOOTSTRAP} and the uncertainties quoted in Equations~\ref{eq:fourier_ASASSN} and \ref{eq:fourier_NEO} show that the sampled coefficients occupy a much smaller region of the parameter space in the $V$-band than the $W$-band. The larger uncertainties in the infrared coefficients are a reflection of the larger intrinsic scatter we noted in the $W$-band period-$\phi_{31}$-[Fe/H] plane (Figure~\ref{fig:FOURIER_COEFFS}) as well as a consequence of the smaller size of the infrared calibrator sample. A detailed analysis of the performance of these relations is provided in Section~\ref{sec:discussion}. Table~\ref{tab:phot_table} lists the derived photometric properties for both the $V$-band and infrared datasets, including the period, $\phi_{31}$ value, and photometric metallicity derived in each band with Equations~\ref{eq:fourier_ASASSN} and \ref{eq:fourier_NEO}. 

\begin{table}
\caption{Derived photometric properties of the RRc sample\label{tab:phot_table}}
\begin{center}
\begin{tabular}{lccccccc}
\tableline
\colhead{Gaia ID} & \colhead{Period\tablenotemark{a}} & \colhead{$\phi_{31}^c$ ($V$)} & \colhead{$\sigma_{\phi_{31}}$ ($V$)} & \colhead{[Fe/H]$_{V}$} & \colhead{$\phi_{31}^s$ ($W$)} & \colhead{$\sigma_{\phi_{31}}$ ($W$)}& \colhead{[Fe/H]$_{W}$}\\
\colhead{(DR3)} & \colhead{(day)} & \colhead{(radian)} & \colhead{(radian)} & \colhead{(dex)} & \colhead{(radian)} & \colhead{(radian)} & \colhead{(dex)}\\
\tableline
\tableline
6914532141197318784 & 0.334879 & 5.002 & 0.017 & $-1.22$ & & & \\
6913110953698726912 & 0.324371 & 3.180 & 0.008 & $-1.66$ & & & \\
6910854717182648448 & 0.323513 & 4.150 & 0.013 & $-1.36$ & & & \\
6897117354482002688 & 0.322392 & 3.418 & 0.007 & $-1.57$ & & & \\
6731321171497007488 & 0.223019 & 2.917 & 0.005 & $-0.71$ & & & \\
6688916306549500800 & 0.339563 & 3.588 & 0.004 & $-1.69$ & 3.244 & 0.041 & $-1.61$\\
6340460627660385920 & 0.285073 & 2.725 & 0.004 & $-1.39$ & & & \\
6340096929829777152 & 0.384128 & 4.077 & 0.011 & $-2.00$ & 2.619 & 0.007 & $-2.19$\\
6307501113055775232 & 0.339337 & 3.149 & 0.009 & $-1.82$ & & & \\
6299550445690958080 & 0.296671 & 3.944 & 0.004 & $-1.15$ & & & \\
\tableline
\end{tabular}  
\end{center}
\tablenotetext{a}{When both $V$-band and infrared (WISE) data is present, the period included was calculated from ASAS-SN ($V$-band) data as the period is usually more accurate due to the higher amplitude and steeper light curve. Period accuracies are quoted to an accuracy on the order of $10^{-6}$ days, corresponding to a readily detectable $\sim$ 1\% shift in phase for a typical 0.32 day period RRc star when phased over the large temporal ($>$ 8 years) baseline of our datasets.
}
\tablecomments{Table~\ref{tab:phot_table} is published in its entirety in machine-readable format. A portion is shown here for guidance regarding its form and content.}
\end{table}

\section{Discussion} \label{sec:discussion}

\subsection{Comparison with Globular Clusters Metallicity} \label{ssec:clusters}

In order to test the $V$-band relation obtained in Section~\ref{ssec:planefit} on an independent sample, we selected a list of ten GCs with metallicity homogeneously spread between [Fe/H] = $-1.0$ and $-2.3$~dex. The clusters sampled are the same as in \citetalias{Mullen2021} with two additional GCs: Reticulum (a Large Magellanic Cloud GC) and NGC~6171. Photometry of the clusters comes from \citealt{Piersimoni2002} (NGC~3201), \citealt{2013AJ....145..160K} (Reticulum), \citetalias{{2014IAUS..301..461M}} (NGC~6171), and the homogeneous data set of P. B. Stetson\footnote{\url{https://www.canfar.net/storage/list/STETSON/homogeneous/Latest_photometry_for_targets_with_at_least_BVI}} (hereafter PBS) for the remaining majority.

\begin{table}[!h]
\caption{Globular Clusters}
\label{tab:globulars}
\begin{center}
\begin{tabular}{lccccc}
\tableline
Clusters & [Fe/H]$_{C09}$ & RRc stars & [Fe/H]$_{RRc}$ & RRab Stars & [Fe/H]$_{RRab}$\\ 
\tableline
NGC 7078 (M15) & $-2.33\pm0.02$ & 36 & $-2.29\pm0.06$ & 64 & $-2.25\pm0.04$ \\ 
NGC 4590 (M68) & $-2.27\pm0.04$ & 9 & $-2.23\pm0.13$ & 13 & $-2.14\pm0.11$ \\ 
NGC 4833 & $-1.89\pm0.05$ & 3 & $-1.92\pm0.40$ & 11 & $-1.95\pm0.11$ \\ 
NGC 5286 & $-1.70\pm0.07$ & 8 & $-1.69\pm0.21$ & 25 & $-1.76\pm0.07$ \\
Reticulum & $-1.67\pm0.12$ & 4 & $-1.74\pm0.05$ & 22 & $-1.58\pm0.03$ \\
NGC 3201 & $-1.51\pm0.02$ & 2 & $-1.64\pm0.40$ & 50 & $-1.38\pm0.08$ \\
NGC 5272 (M3) & $-1.50\pm0.05$ & 28 & $-1.54\pm0.12$ & 175 & $-1.39\pm0.08$ \\ 
NGC 5904 (M5) & $-1.33\pm0.02$ & 21 & $-1.37\pm0.08$ & 67 & $-1.41\pm0.07$ \\ 
NGC 6362 & $-1.07\pm0.05$ & 13 & $-1.22\pm0.08$ & 18 & $-1.16\pm0.06$ \\ 
NGC 6171 (M107) & $-1.03\pm0.02$ & 8 & $-1.20\pm0.04$ & 15 & $-0.81\pm0.13$ \\ 
\tableline
\end{tabular}  
\end{center}

\end{table}

The general properties of the clusters are listed in Table~\ref{tab:globulars}, which includes their spectroscopic metallicities (in the scale of \citetalias{Carretta2009}) and the number of \rrl{} with good-quality well-sampled light curves available in each cluster. The spectroscopic metallicity of the Galactic GCs are obtained from \citetalias{Carretta2009}, while the metallicity of Reticulum is from \citet{2004MNRAS.355..504M} and converted from the [Fe/H] scale of \citet[][ZW84]{ZinnWest1984} to \citetalias{Carretta2009}. 
A Fourier decomposition was performed on each light curve to obtain their $\phi_{31}$ parameters, with the exception of the stars in NGC~6171 and NGC~3201 for which the Fourier parameters were taken directly from their respective photometric catalogs. The period-$\phi_{31}$-[Fe/H] relations for RRc (Equation~\ref{eq:fourier_ASASSN}) and RRab (Equation 6 from \citetalias{Mullen2021} with [Fe/H]'s shifted to the scale of \citetalias{Carretta2009}) were then applied to estimate the metallicity of each \rrl{} star in the clusters. Note, no metallicity scale corrections are needed between that of \citetalias{Carretta2009} and the RRc relations presented in this work. The average [Fe/H] abundance of the RRc and RRab variables in each cluster, with their standard deviation, are listed in the 4th and 6th columns of Table~\ref{tab:globulars}, respectively.

\begin{figure}[h!]
    \centering
    \includegraphics[scale=0.6]{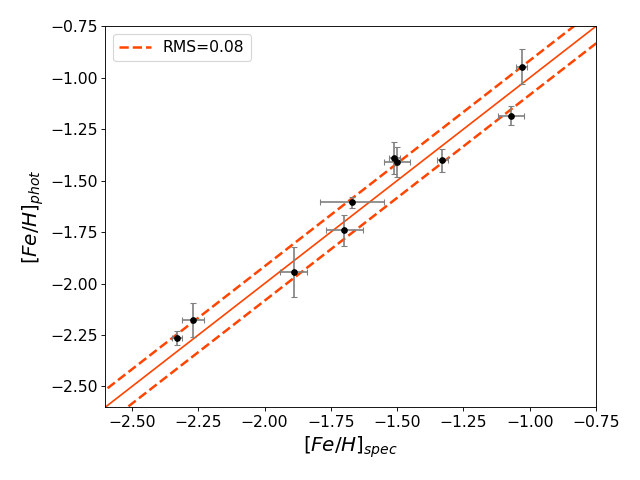}
    \caption{[Fe/H]$_{\rm{spec}}$ versus [Fe/H]$_{\rm{phot}}$ obtained by using Equation~\ref{eq:fourier_ASASSN} (this work) for RRc variables, and Equation 6 of \citetalias{Mullen2021} (shifted to the scale of \citetalias{Carretta2009}) for the RRab, in a sample of 10 GCs.  The solid orange line is the 1-1 relation, while the dashed orange lines show the standard deviation ($\pm$0.08~dex). Error bars correspond to the uncertainties in spectroscopic metallicity (from \citetalias{Carretta2009} and \citealt{2004MNRAS.355..504M}) and the statistical error in the photometric metallicity calculated as standard deviation of the metallicity of the individual \rrl{} in the cluster.
    }
    \label{fig:feh_comparison_GCs}
\end{figure}

Figure~\ref{fig:feh_comparison_GCs} shows the spectroscopic [Fe/H] versus the mean photometric [Fe/H] values, calculated for each GC with our relations. Each RRab and RRc stellar photometric metallicity measurement contributes equal weight to the mean photometric [Fe/H] for a given GC. The figure, as well as the individual values listed in Table~\ref{tab:globulars}, demonstrates the good performance of our formul\ae{} (including the new relations for RRc variables) to provide reliable [Fe/H] abundances. The combination of the photometric RRab metallicities (from \citetalias{Mullen2021}) with the new photometric RRc metallicities is in good agreement with the spectroscopic metallicities of the clusters: overall $\pm$0.08~dex, well within the respective uncertainties.

\subsection{Comparison with Literature Relations}\label{ssec:validation}

\begin{figure}[!h]
        \centering
        \includegraphics[width=\textwidth]{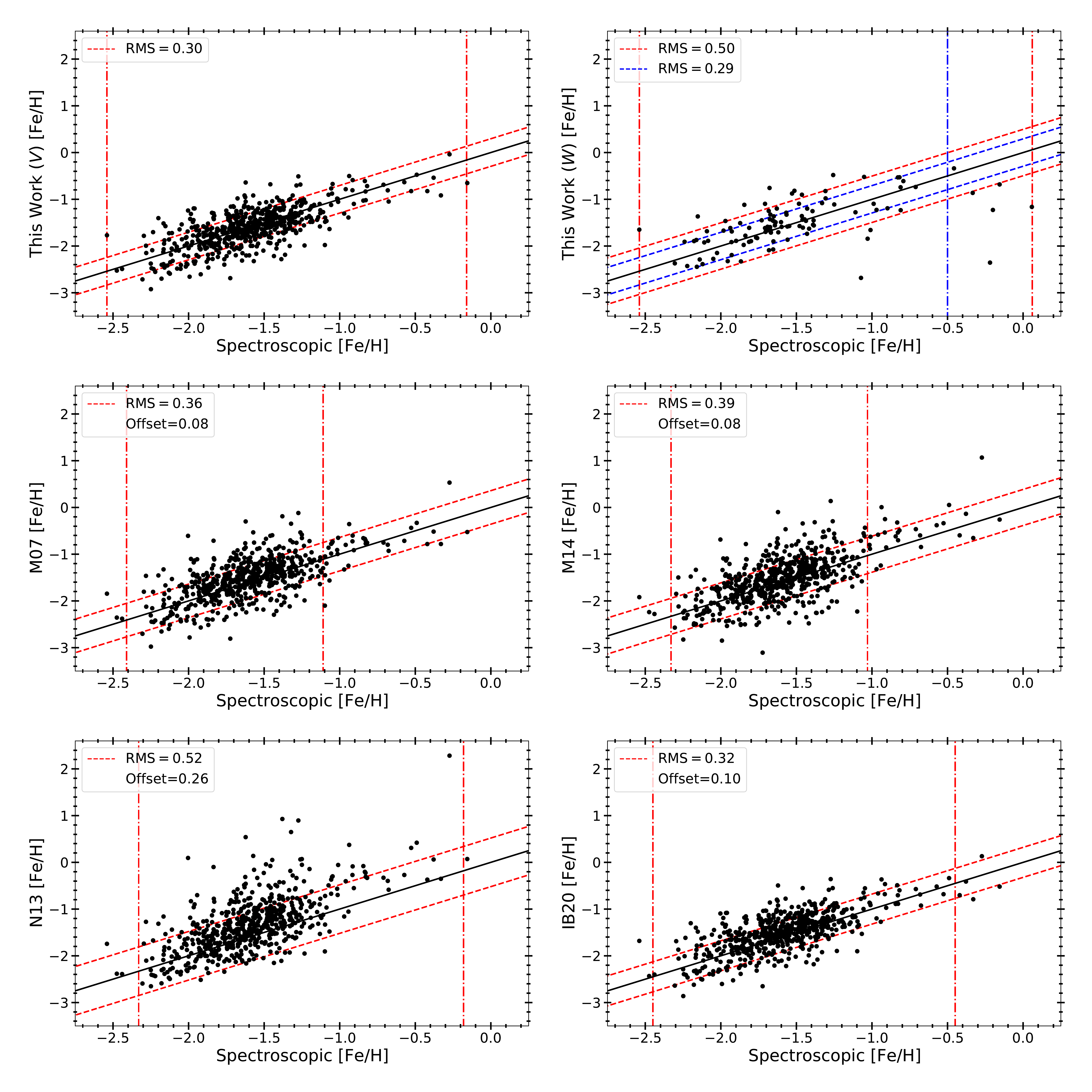}
        \caption{Comparison of the spectroscopic [Fe/H] abundances of our calibration sample and those derived with a variety of photometric metallicities, such as the ones presented in Section~\ref{ssec:processing} (top row), \citetalias{2007MNRAS.374.1421M} and \citetalias{2014IAUS..301..461M} (middle row), \citetalias{2013ApJ...773..181N} and \citetalias{2020arXiv200802280I} (bottom row). The diagonal dashed lines represent the RMS dispersion about the ideal one-to-one relation. The metallicity range used to calibrate each relation is indicated by the two vertical red dash-dotted lines. An additional RMS is shown for WISE band relation (top-right panel, blue dashed lines), calculated using only the stars with $\textrm{[Fe/H]} \la -0.5$ (vertical blue dash-dotted line).}
        \label{fig:SpecRelations}
\end{figure}

In this section, we compare photometric metallicities derived using a variety of period-$\phi_{31}$-[Fe/H] relations, including the ones found in this work (Equations~\ref{eq:fourier_ASASSN} and \ref{eq:fourier_NEO}), as well as the fits provided by \citetalias{2007MNRAS.374.1421M}, \citetalias{2014IAUS..301..461M}, \citetalias{2013ApJ...773..181N}, and \citetalias{2020arXiv200802280I}. Figure~\ref{fig:SpecRelations} shows the photometric metallicities for all the RRc in our calibration sample, plotted as a function of their spectroscopic [Fe/H] abundances. The RMS scatter of each relation, calculated with respect to the ideal one-to-one relationship over the entire spectroscopic metallicity range within which each relation has been calibrated, is indicated in each case.

The top row shows the analysis of our fits in the optical (ASAS-SN, left) and infrared (WISE, right) bands. Since we are directly comparing to the same sample used to derive these relations, the figure allows us to check for remaining trends in the residuals with respect to the ideal one-to-one relation to ensure that our relations provide a consistent estimate of each stars' [Fe/H] over the entire metallicity range of our calibration sample. For the case of the $V$ -band ASAS-SN dataset, we show nearly symmetric residuals (with an $\textrm{RMS} \approx 0.30$~dex) over the entire range of metallicities with only a minor deviation at higher metallicities, still well within the error of our relation. The WISE bands' residuals, however, show some larger divergence at high metallicity ($\textrm{[Fe/H]} \ga -0.5$), where Equation~\ref{eq:fourier_NEO} systematically under-predicts the spectroscopic metallicities. The overall residual RMS is also significantly larger (0.50~dex) than the one we obtained in the $V$-band. Due to the small number of the high metallicity RRc stars, we cannot determine if this is due to non-linearity of the period-$\phi_{31}$-[Fe/H] in this metallicity regime, or rather a reflection of less accurate spectroscopic metallicities for the stars in our calibration sample that approach solar [Fe/H] abundance. Indeed, \citet{2021ApJ...914...10C} found that the \rrl{} that have been used to calibrate the $\Delta S$ method in \citet{2021ApJ...908...20C}, which in turn we use as the basis for our own relation, show a broad dispersion in $\alpha$-elements measured through high resolution spectroscopy. Since the $\Delta S$ method is based on the strength of a Ca line, while our Fourier-$\phi_{31}$-metallicity method aims to measure [Fe/H] metallicities, a spread in $\alpha$-element abundances can potentially lead to the observed deviations at the ends of the metallicity scale. If we restrict our WISE bands analysis to $\textrm{[Fe/H]} \la -0.5$, we obtain a best fit relation that is virtually indistinguishable from Equation~\ref{eq:fourier_NEO} (because of the much larger number of low metallicity RRc, and due to the robustness of our bootstrap fit), but with residual RMS equal to 0.29~dex, nearly identical to the one we found for the $V$-band (calculated over the entire metallicity range). By directly comparing the V-band (Equation~\ref{eq:fourier_ASASSN}) and infrared (Equation~\ref{eq:fourier_NEO}) photometric metallicities for the RRc in the joint sample, we note that the two sets of metallicities are consistent with each other, within their respective uncertainties.

The middle-left panel shows the residuals found with the higher-order nonlinear period-$\phi_{31}$-[Fe/H] relation of \citetalias{2007MNRAS.374.1421M} (their equation 4). Note that \citetalias{2007MNRAS.374.1421M} published two equations: one based on the [Fe/H] scale of \citetalias{ZinnWest1984} and one based on that of \citet{1997A&AS..121...95C}. We choose to analyze the latter due to its smaller quoted dispersion and simpler functional form. Their relation was based on 106 RRc stars from 12 GCs with $V$-band Fourier parameters gathered from heterogeneous publications. In order to consistently compare with the HR+$\Delta$S spectroscopic metallicities, we converted the metallicities derived from the \citetalias{2007MNRAS.374.1421M} relation to that of \citetalias{Carretta2009} using the relation $\rm{[Fe/H]}_{C09}=(1.137 \pm 0.060)\rm{[Fe/H]}_{CG97}-0.003$ (provided by \citetalias{Carretta2009}). A few years later, \citetalias{2014IAUS..301..461M} published an updated version of their $V$-band \citetalias{2007MNRAS.374.1421M} fit, using the same base of RRc extended to include a total of 163 stars gathered from 19 GCs, with a metallicity scale updated to that of \citetalias{Carretta2009}. Although they found a slightly different nonlinear period-$\phi_{31}$-[Fe/H] functional form, the metallicities predicted by  \citetalias{2014IAUS..301..461M} (center-right panel in Figure~\ref{fig:SpecRelations}) appear to be roughly the same as those provided by \citetalias{2007MNRAS.374.1421M}. Both relations do not appear to have a trend or bias with respect to the spectroscopic metallicities, but they do have a larger RMS (0.36 and 0.39~dex. respectively) than our ASAS-SN relation.

\citetalias{2013ApJ...773..181N} provided a nonlinear period-$\phi_{31}$-[Fe/H] for the Kepler Space Telescope's $Kp$ band (their Equation 4). Due to the small field-of-view (relatively to large-area surveys) of Kepler's primary mission, this relation is calibrated by augmenting the sample of 3 RRc obtained in Kepler's field with the Fourier parameters of 98 GC RRc from \citetalias{2007MNRAS.374.1421M}, converted to the $Kp$ system using the relation $\phi_{31}(V)=\phi_{31}(Kp)-(0.151\pm0.026)$, from \citealt{2011MNRAS.417.1022N}. We use this same relation to convert the $V$-band $\phi_{31}$ parameters of our spectroscopic sample to the $Kp$ photometric system in order to generate the bottom-left panel of Figure~\ref{fig:SpecRelations}. The plot shows that the \citetalias{2013ApJ...773..181N} largely predicts the [Fe/H] metallicities without noticeable trends even outside their [Fe/H] calibration range, albeit with a small negative bias (the average metallicity predicted by \citetalias{2013ApJ...773..181N} is 0.26~dex lower than the spectroscopic values), and a larger scatter ($\textrm{RMS} = 0.52$~dex) than all other relations here evaluated.

Finally, \citetalias{2020arXiv200802280I} used a sample of 50 GC RRc stars extracted from the Gaia DR2 database to derive a bi-linear period-$\phi_{31}$-[Fe/H] relation (Equation 4 in their paper) with the same functional form of our Equations~\ref{eq:fourier_ASASSN} and \ref{eq:fourier_NEO}. In order to apply this relation to our dataset, we had to perform two transformations: (1) we converted the $V$-band $\phi_{31}$ value of our ASAS-SN sample to the $G$-band system, using the relation $\phi_{31}(G)=(0.104\pm0.020)+(1.000\pm0.008) \, \phi_{31}(V)$ from \citet{2016A&A...595A.133C}, and (2) we transformed the metallicity provided by the \citetalias{2020arXiv200802280I} relation (in the \citetalias{ZinnWest1984} scale) into the Carretta scale using the $\rm{[Fe/H]_{C09}}=1.105\rm{[Fe/H]_{ZW84}}+0.160$ relation from \citetalias{Carretta2009}. The results are shown in the bottom-right panel of Figure~\ref{fig:SpecRelations}. Again, no obvious trend or offset is found, and the RMS dispersion (0.32~dex) is among the smallest of the relations assessed in this Section, comparable to the RMS we measure in the $V$-band with our Equation~\ref{eq:fourier_ASASSN}.

Overall, \citetalias{2007MNRAS.374.1421M}, \citetalias{2014IAUS..301..461M}, and \citetalias{2013ApJ...773..181N} share the majority of their photometric calibration dataset but differ in their functional form by adding slightly different combinations of nonlinear terms. Figure~\ref{fig:SpecRelations} shows that, with respect to optical \emph{linear} relations (\citetalias{2020arXiv200802280I} and Equation~\ref{eq:fourier_ASASSN} in this work), these non-linear relations tend to have larger residual dispersion for our large sample of spectroscopic metallicities. With the exception of the infrared WISE band (where we observe a possible departure at solar metallicities), our dataset does not support the need for the addition of non-linear terms in photometric metallicities relations based on Fourier parameters of RRc optical lightcurves.

\subsection{Comparison with Sculptor dSph Metallicity} \label{ssec:sculptor}

In this section, we test our $V$-band period-$\phi_{31}$-[Fe/H] relations with \rrl{} stars in the Milky Way's dSph satellite Sculptor. We compare our photometric Fourier metallicities to those derived by \citet{2016MNRAS.461L..41M} (hereafter \citetalias{2016MNRAS.461L..41M}) by inverting the theoretical PLZ relation from \citet{2015ApJ...808...50M}. Being a relatively nearby local group galaxy ($\mu_0 = 19.62$~mag; \citealt{2015MNRAS.454.1509M}), Sculptor has been extensively studied as a probe for galaxy evolution, and detailed studies are available of its variable star content with photometry stretching back over two decades (for a thorough review see \citealt{2016MNRAS.462.4349M}, hereafter \citetalias{2016MNRAS.462.4349M}). We specifically chose Sculptor as a test case for our Fourier-metallicity relations as this dwarf galaxy has been shown to have an early history of chemical enrichment, resulting in an older stellar population (including \rrl{} stars) with a broad range of metallicity ($\ga$1 dex, see \citealt{2005MNRAS.363..734C,2015MNRAS.454.1509M,2016MNRAS.461L..41M}). By studying Sculptor, we show that the relations provided in this work are widely applicable to complex stellar populations, beyond \rrl{}s in the field or GCs.

Sculptor photometry is available from the PBS database in both the $I$-band (utilized by \citetalias{2016MNRAS.461L..41M} in their work) and the $V$-band. We refer to \citetalias{2016MNRAS.462.4349M} for the exact details of the observing runs, bands observed, instruments and telescopes used to collect the Sculptor photometry in PBS. Out of 536 known \rrl{}s  (289 RRab, 197 RRc, and 50 RRd; \citetalias{2016MNRAS.462.4349M}), \citetalias{2016MNRAS.461L..41M} derived photometric metallicities for 276 RRab and 195 RRc stars. However, only 277 of these stars (126 RRab and 148 RRc)have good quality $V$-band light curves (according to the criteria outlines in \citetalias{Mullen2021}, Section 3). For these stars we have extracted their $V$-band Fourier parameters, following the procedures described in Section~\ref{ssec:processing}. We have then estimated their photometric [Fe/H] abundances, using our Fourier-metallicity relations for RRab (Equation~6 from \citetalias{Mullen2021} with [Fe/H]'s shifted to the scale of \citetalias{Carretta2009}) and RRc (Equation~3 from section~\ref{ssec:planefit}) stars respectively.

\begin{figure}[h!]
    \centering
    \includegraphics[width=\textwidth]{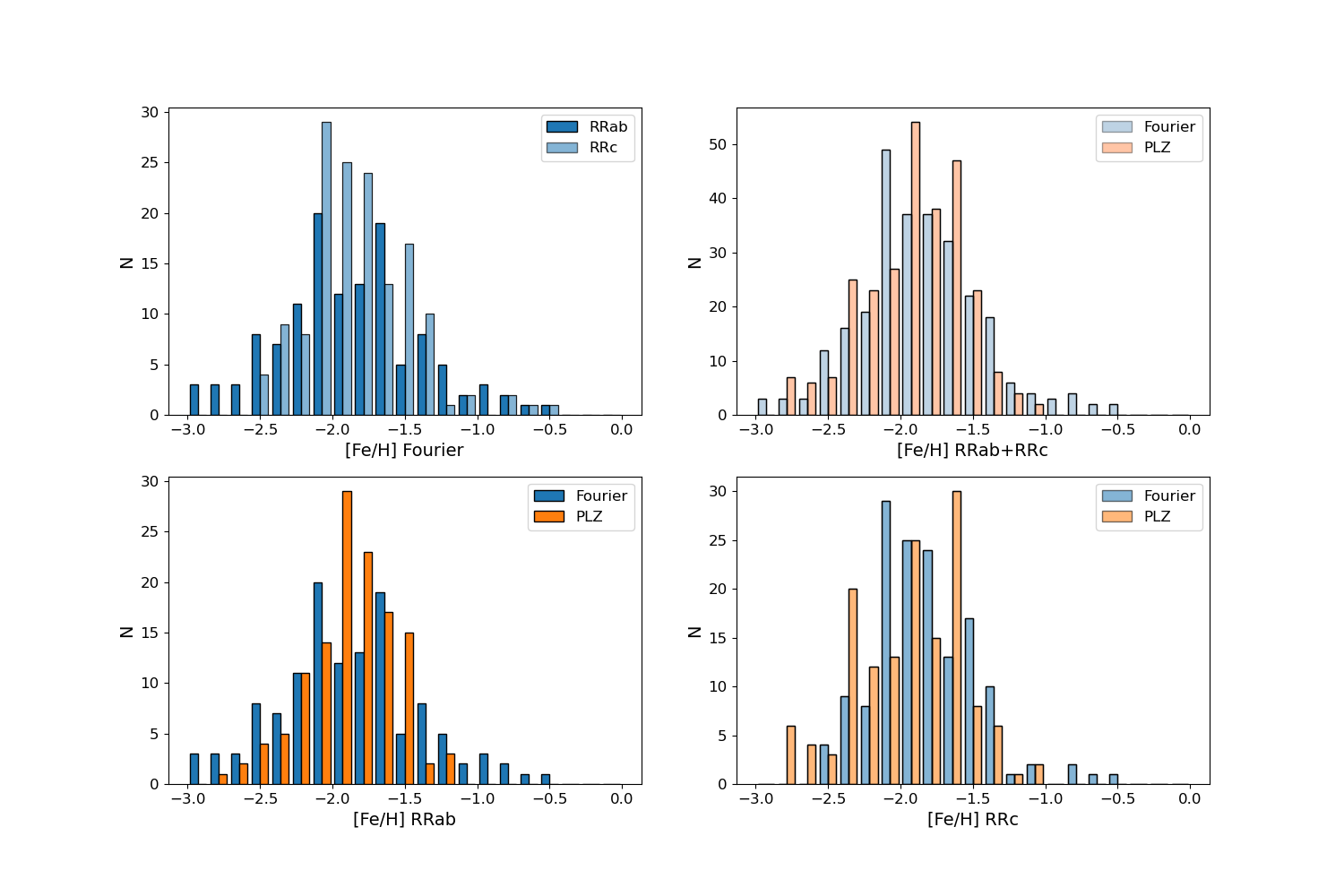}
    \caption{Comparison of derived [Fe/H] distributions of Sculptor dSph variables. \emph{Top Left}: Histogram of the [Fe/H] obtained by using the Fourier $V$-band $\phi_{31}$ applied to Equation~\ref{eq:fourier_ASASSN} (this work) for RRc variables, and Equation 6 of \citetalias{Mullen2021} (shifted to the scale of \citetalias{Carretta2009}) for the RRab.
    Remaining panels compare the Fourier $\phi_{31}$ derived [Fe/H] to those derived in \citetalias{2016MNRAS.461L..41M} (through inversion of an $I$-band PLZ relation, see text) for the entire RRab+RRc sample (\emph{Top Right}), and solely RRab or RRc variables (\emph{Bottom Left} and \emph{Bottom Right}, respectively).}
    \label{fig:sculptor_distrib}
\end{figure}

The results of this analysis are shown in Figure~\ref{fig:sculptor_distrib}. The top-left panel shows the Fourier metallicity distributions of Sculptor's RRab and RRc stars. The two histograms have a similar mean ($-1.88$~dex and $-1.83$~dex for the RRab and RRc, respectively) and dispersion ($\sigma_{RRab}=0.48$~dex and $\sigma_{RRc}=0.36$~dex). The remaining panels of Figure~\ref{fig:sculptor_distrib} compare the metallicity distribution obtained with our Fourier method with the [Fe/H] abundance derived via PLZ inversion by \citetalias{2016MNRAS.461L..41M}. We find a remarkable agreement between the metallicities derived with these two methods. In the overall sample (RRab and RRc together, top-right panel) we measured a mean metallicity $\textrm{[Fe/H]} \simeq -1.85$~dex, matching the value of $\textrm{[Fe/H]} \simeq -1.90$~dex found by \citetalias{2016MNRAS.461L..41M}. The dispersion and shape of the two [Fe/H] distributions is also very similar, with both methods suggesting a spread of metallicities in the Sculptor \rrl{} population of $\sim 2$~dex. Note that while the Fourier metallicities derived in this work suggest the presence of a small high-metallicity tail (with a few stars having [Fe/H] between $-1.0$ and $-0.5$~dex, not seen in \citetalias{2016MNRAS.461L..41M}), this excess may be an artifact originating in the calibration of our relation, as noted in section~\ref{ssec:validation}. Comparison of the Fourier and PLZ metallicities derived for RRab and RRc separately (bottom row in figure~\ref{fig:sculptor_distrib}) leads to similar conclusions.

These results are also consistent with Sculptor's spectroscopic metallicities. \citet{2005MNRAS.363..734C} found a [Fe/H] peaking at $\sim-1.8$~dex via the $\Delta$S spectroscopic method applied to a sample of 107 \rrl{}. Achieving such a consistent mean metallicity and distribution validates both of these photometric approaches, especially since the two photometric relations are completely independent of one another both in methodology (shape of the light curve as opposed to inversion of the PLZ relation) and in the dataset used ($V$ band vs. $I$ band, and light curves with different sampling and coverage).

\section{Conclusions} \label{sec:conclusions}

In this work, we provide new relations to derive photometric metallicities based on the $\phi_{31}$ Fourier parameter of optical and, for the first time, infrared light curves of RRc variables. Our relations are calibrated using a large sample ($\sim 4 \times$ larger than anything used prior at optical wavelengths) of field RRc variables for which homogeneous spectroscopic abundances are available and cover a broad range of metallicities ($-2.5 \la \textrm{[Fe/H]}\la 0.0$) derived from HR spectra and the $\Delta S$ method, using techniques developed by \citetalias{2021ApJ...908...20C}. The photometric time series of our calibration stars were extracted from the ASAS-SN ($V$ band) and the WISE (NEOWISE extension, $W1$ and $W2$ bands) surveys, providing well-sampled light curves that allow for reliable Fourier expansions.

Comparisons with other optical photometric metallicity relations for RRc variables show that our formula provides reliable [Fe/H] abundances without noticeable trends over the entire metallicity range found in the Milky Way halo. Our $V$-band relation (Equation~\ref{eq:fourier_ASASSN}) reproduces spectroscopic metallicities with a residual standard deviation of $\simeq 0.30$~dex, smaller than the higher-order relations found in the literature. We tested our $V$-band relation on \rrls{} in GCs and shown that we can accurately estimate the average clusters' metallicity with an overall accuracy of $\sim 0.08$~dex. Additionally, we have shown this relation can reproduce the [Fe/H] distribution of systems with a more complicated history of chemical enrichment, such as the Sculptor dSph, consistent with the predictions of both spectroscopy and other photometric relations.

For the first time, We have also obtained a mid-infrared period-$\phi_{31}$-[Fe/H] relation in the WISE $W1$ and $W2$ bands (Equation~\ref{eq:fourier_NEO}). Despite having a calibration sample five times smaller than the $V$-band sample, our mid-infrared relation has similar accuracy ($\simeq 0.29$~dex) in the low and moderate metallicity range ($\textrm{[Fe/H]} \la -0.5$~dex). In the high metallicity range, our relation appears to under-predict the [Fe/H] abundance expected from spectroscopy; further analysis (relying on a larger sample of solar-metallicity RRc calibrators) is needed to understand the root cause of this deviation.

This work complements the analysis we presented in \citetalias{Mullen2021}, where we derived similar relations for field RRab variables using the same techniques described here (and a calibration catalog with spectroscopic metallicities derived with the same methods). Section~\ref{ssec:clusters} shows that, for stellar populations where both RRab and RRc variables are found (e.g., many Galactic GCs), combining spectroscopic metallicities from \rrl{} in both pulsation modes further improves the [Fe/H] reliability, with an accuracy of the population average metallicity approaching high-resolution spectroscopic measurements.

Whether in the mid-infrared or optical, for the RRab or RRc, the relations presented here will be crucial to facilitate the quick determination of reliable \rrl{} metallicities for the many upcoming wide-area time-domain surveys and ELTs (such as LSST at the Vera C. Rubin Observatory). Our mid-infrared relation will allow future telescopes (such as JWST and the Roman telescope) to reach RRLs across the Local Group of galaxies, where spectral observations will not be feasible. Finally, by providing a method to obtain reliable metallicities of individual \rrl{}, \citetalias{Mullen2021} and this work will be crucial in determining accurate distances with PLZ and PWZ relations.

\acknowledgments
This publication makes use of data products from WISE, which is a joint project of the University of California, Los Angeles, and the Jet Propulsion Laboratory (JPL)/California Institute of Technology (Caltech), funded by the National Aeronautics and Space Administration (NASA), and from NEOWISE, which is a JPL/Caltech project funded by NASA’s Planetary Science Division. 

This publication also makes use of data products from the ASAS-SN project, which has their telescopes hosted by Las Cumbres Observatory. ASAS-SN is supported by the Gordon and Betty Moore Foundation through grant GBMF5490 and the NSF by grants AST-151592 and AST-1908570. Development of ASAS-SN has been supported by Peking University, Mt. Cuba Astronomical Foundation, Ohio State University Center for Cosmology and AstroParticle Physics, the Chinese Academy of Sciences South America Center for Astronomy (CASSACA), the Villum Foundation, and George Skestos.

This work has made use of data from the European Space Agency (ESA) mission
{\it Gaia} (\url{https://www.cosmos.esa.int/gaia}), processed by the {\it Gaia}
Data Processing and Analysis Consortium (DPAC,
\url{https://www.cosmos.esa.int/web/gaia/dpac/consortium}). Funding for the DPAC
has been provided by national institutions, in particular the institutions
participating in the {\it Gaia} Multilateral Agreement.

Guoshoujing Telescope (the Large Sky Area Multi-Object Fiber Spectroscopic Telescope LAMOST) is a National Major Scientific Project built by the Chinese Academy of Sciences. Funding for the project has been provided by the National Development and Reform Commission. LAMOST is operated and managed by the National Astronomical Observatories, Chinese Academy of Sciences.

    Funding for the SDSS and SDSS-II has been provided by the Alfred P. Sloan Foundation, the Participating Institutions, the National Science Foundation, the U.S. Department of Energy, the National Aeronautics and Space Administration, the Japanese Monbukagakusho, the Max Planck Society, and the Higher Education Funding Council for England. The SDSS Web Site is http://www.sdss.org/. The SDSS is managed by the Astrophysical Research Consortium for the Participating Institutions. The Participating Institutions are the American Museum of Natural History, Astrophysical Institute Potsdam, University of Basel, University of Cambridge, Case Western Reserve University, University of Chicago, Drexel University, Fermilab, the Institute for Advanced Study, the Japan Participation Group, Johns Hopkins University, the Joint Institute for Nuclear Astrophysics, the Kavli Institute for Particle Astrophysics and Cosmology, the Korean Scientist Group, the Chinese Academy of Sciences (LAMOST), Los Alamos National Laboratory, the Max-Planck-Institute for Astronomy (MPIA), the Max-Planck-Institute for Astrophysics (MPA), New Mexico State University, Ohio State University, University of Pittsburgh, University of Portsmouth, Princeton University, the United States Naval Observatory, and the University of Washington.

M. Marengo and J. P. Mullen are supported by the National Science Foundation under Grant No. AST-1714534. 

%

\vspace{5mm}
\facilities{WISE, ASAS-SN, \textit{Gaia}, LAMOST, SDSS-SEGUE}


\software{Astropy \citep{2013A&A...558A..33A}, SciPy \citep{2020SciPy-NMeth}, scikit-learn \citep{scikit-learn}}





\bibliographystyle{aasjournal}



\end{document}